\documentclass[fleqn,12pt,twoside]{article}
\usepackage{espcrc1}

\usepackage{graphicx}
\newcommand{\beq}{\begin{equation}}
\newcommand{\eeq}[1]{\label{#1}\end{equation}}

\def\be{\begin{equation}}
\def\ee{\end{equation}}
\def\bea{\begin{eqnarray}}
\def\eea{\end{eqnarray}}

\newcommand{\AmS}{{\protect\the\textfont2
  A\kern-.1667em\lower.5ex\hbox{M}\kern-.125emS}}

\title{Modified Boltzmann Transport Equation}

\author{V.K. Magas\address{Departamento de F\'{\i}sica Te\'orica and IFIC, Centro Mixto,\\ 
     Institutos de Investigaci\'on de Paterna - Universidad de Valencia-CSIC\\ 
     Apdo. correos 22085, 46071, Valencia, Spain},
        L.P. Csernai\address[Bergen]{Section for Theoretical and Computational Physics,\\ 
     and Bergen Computational Physics Lab, BCCS-Unifob,\\
     University of Bergen, Allegaten 55, 5007 Bergen, Norway},
     E. Moln\'ar\addressmark[Bergen],
     A. Nyiri\addressmark[Bergen], 
     and
     K. Tamosiunas\addressmark[Bergen]  
}
\begin{document}

% typeset front matter
\maketitle

\begin{abstract}
Recently several works have appeared in the literature in which authors try to describe 
Freeze Out (FO) in energetic heavy ion collisions based on the Boltzmann Transport Equation (BTE). 
The aim of this work is to point out the limitations of the BTE, when applied for the modeling of 
FO or other very fast process, 
and to propose the way how the BTE approach can be generalized for such a processes. 
\end{abstract}

\vspace{0.5cm}

The Freeze Out (FO) is an important phase of dynamical reactions.
The FO is a kinetic process and thus the phase-space (PS) distribution of post FO
particles can be obtained from kinetic FO calculations. 
The connection of the kinetic description of this process and
the Boltzmann Transport Equation (BTE) raised considerable attention recently
\cite{E,B}.
One would think that FO can be
handled perfectly by using the BTE, which
may describe both equilibrium and non-equilibrium processes 
in a 4-dimensional space-time (ST) volume element like
FO layer.  Our
aim is to analyse the situation and point out the physical causes, which
limit the applicability of the BTE for describing FO and some other processes \cite{MBTE}.

FO is usually assumed to happen on sharp 3-dimensional
ST hypersurfaces. However, the FO-fronts or FO-layers are not 
necessarily narrow, but typically they have
a characteristic direction or normal, $d^3\sigma^\mu$,  
which can be time-like or space-like. It is more
realistic to assume a continuous, 4-volume FO in a layer (or domain)
of the ST. At the inside boundary of this layer
there are only interacting particles, while at the outside boundary hypersurface
all particles are frozen out and no interacting particles remain.
Assuming that the boundaries of this layer are
approximately parallel and neglecting derivatives in all directions other than  
$d^3\sigma^\mu$, we can consider FO process as 1-dimensional \cite{MBTE,FO}.

We can derive the BTE from the conservation
of charges in a ST domain, $\Delta^4x$, assuming the
standard conditions, see for example \cite{LasBook}:\\
(i) only binary collisions are considered,\\
(ii) we assume "molecular chaos", i.e. that the number of
     binary collisions at position $x$ is proportional to
     $f(x,p_1) \times f(x,p_2)$, \\
(iii) $f(x,p)$ is a smoothly varying function compared to the
mean free path (m.f.p.).\\
 We have to
take into account that particles can scatter into the phase space (PS) volume
element around $p$, or can scatter out from this volume element, 
described by Gain- and Loss- collision terms in the BTE.
In these terms
we consider elementary binary collisions where in the initial state
two particles collide with momenta
$p_1$ and $p_2$
into a final state of two particles with momenta
$p_3$ and $p_4$. In case of the Gain term the particle described by the
BTE, with momentum $p$ (without an index), is one from the two final
state particles, while in case of the loss term this particle is
one of the initial state particles. This is indicated by the
indexes of the invariant transition rate, $W_{34}^{p2}$ and $W_{p2}^{34}$ 
correspondingly \cite{LasBook}.
We integrate over the momenta of the other three particles
participating in this binary collision. We use the following notation
for the PS integrals:
$$
{D}_{123} \equiv
\frac{d^3 p_1}{p_1^0}
\frac{d^3 p_2}{p_2^0}
\frac{d^3 p_3}{p_3^0} \,.
$$
So finally we can write BTE in the following way:
\beq
p^\mu \partial_\mu f(x,p) 
=\frac{1}{2}\int {D}_{234} W_{34}^{p2} \left[f(x,p_3) f(x,p_4)-f(x,p)   f(x,p_2)\right]\ .
\eeq{BTE}
Here we want to underline the important symmetry properties of the invariant transition rate, namely
$W_{12}^{34}=W_{21}^{34}=W_{34}^{12}$. This transition rate can be expressed  as $W_{12}^{34}=s 
\sigma_{diff} \delta(p_1+p_2-p_3-p_4)$, where $\delta$-function enforces energy-momentum conservation,
$s=(p_1+p_2)^2$ and $\sigma_{diff}$ is a differential cross section, here we will assume that it is momentum
independent (so the total cross section $\sigma=4\pi \sigma_{diff}$). 

However, the usual structure of the collision terms in the
BTE is not adequate for describing rapid process in a layer with a thickness 
comparable with the m.f.p. If we assume the existence of such a
layer this immediately
contradicts assumption (iii):
the change is not negligible in the direction of
$d^3 \sigma^\mu$ (normal to the layer). The assumption of
"molecular chaos" (ii) is also violated in such a  process, because
number of collisions is not proportional to
$f(x,p_1) \times f(x,p_2)$, but it is delocalized
in the normal direction with
$f(x_1,p_1) \times f(x_2,p_2)$, where $x_k$ is the origin of colliding particles, i.e., the ST point
where the colliding particles were colliding last.

Based on the above considerations, one might conclude that the changes
of the distribution function are mediated by the transfer of particles, and
consequently only slowly propagating changes are possible. If the FO
layer propagates slowly, then its normal, $d^3\sigma^\mu$, must always 
be space-like.
This was a common misconception, where all "superluminous"
shock, detonation, deflagration fronts or discontinuities, and FO were considered
unphysical based on early studies \cite{Taub}.
However, later it was shown that discontinuous changes may
happen simultaneously in spatially neighbouring points, i.e. the
normal of the discontinuity-hypersurface can be time-like \cite{Cs87}.
This applies to the FO process also. Thus, the direction of characteristic
or dominant change,  $d^3\sigma^\mu$, may be both space-like and time-like
in the FO process.

From the all processes mentioned above (i.e. shocks, detonations, 
deflagrations etc.) 
the FO is the most special one, because the number of interacting 
particles is constantly 
decreasing as the FO proceeds, correspondingly the m.f.p. 
is increasing. In fact, it reaches infinity
when the FO is completed. 
This means that we 
can not make the FO in finite layer of
any thickness smooth enough to be modeled with the BTE! 
It is also obvious that if FO has some 
characteristic length scale,
it is not proportional to the m.f.p., because the m.f.p. increases 
as the density of interacting component becomes smaller, 
while the FO becomes faster in this limit, so its 
characteristic scale should decrease.

To describe that the PS distributions which change rapidly, faster than the m.f.p., we suggest a 
Modified Boltzmann Transport Equation (MBTE):
\beq
p^\mu \partial_\mu f(x,p) =
 \frac{1}{2}\int {D}_{234}\, W_{34}^{p2}\,  \left[\overline{f(x,p_3)}^{\ x}\ \overline{f(x,p_4)}^{\ x}\  
- \overline{f(x,p)}^{\ x}\ \overline{f(x,p_2)}^{\ x}\right]\,, \\
\eeq{mbte}
where $\overline{f(x,p_i)}^{\ x}$ is an average over all possible origins of the particle in the 
backward lightcone of
the ST point $x=(t,\vec{x})$:
\beq
 \overline{f(x,p)}^{\ x}=\frac{\int_{t_0}^{t} dt_1 \int d^3x_1 \delta^3(\vec{x} - \vec{x}_1 - \vec{v} (t-t_1)) f(x_1,p) 
 e^{-\int_{t_1}^{t} dt_2 \int d^3x_2 \sigma n(x_2) v \delta^3(\vec{x}_2 - \vec{x}_1 - \vec{v} (t_2-t_1))}} 
 {\int_{t_0}^{t} dt_1 \int d^3x_1 \delta^3(\vec{x} - \vec{x}_1 - \vec{v} (t-t_1)) 
 e^{-\int_{t_1}^{t} dt_2 \int d^3x_2 \sigma n(x_2) v \delta^3(\vec{x}_2 - \vec{x}_1 - \vec{v} (t_2-t_1))}}, 
\eeq{aver}
where $\delta^3(\vec{x} - \vec{x}_1 - \vec{v} (t-t_1))$ fixes the ST 
trajectory along which the particles with given momentum can reach the ST point $x$, time 
$t_0$ is given by the
initial or boundary conditions,  
$\vec{v} = \vec{p} / p^0$ ($v=|\vec{v}|$),
and the exponential factor accounts for the probability not to have any
 other collision from the origin point
$x_1$ till $x$. In the arguments of exponents $n(x)$ is the particle density in the calculational frame, 
$n(x)=\int d^3 p f(x,p)$, 
$\sigma$ is the total  scattering cross section. After performing integrations over $d^3x$ with a help of 
$\delta$-functions we can 
write MBTE equation in the form:
\beq
p^\mu \partial_\mu f =
\int \mathcal{D}_{234}^{t_1t_2}
\left[f(t_1,p_3) G(t_1,p_3)   
f(t_2,p_4)  G(t_2,p_4) - 
f(t_1,p) G(t_1,p)   
f(t_2,p_2)  G(t_2,p_2)\right],
\eeq{mbte2}
where
\beq
\mathcal{D}_{234}^{t_1t_2}=\frac{1}{2}\int_{t_0}^{t} dt_1 \int_{t_0}^{t} dt_2 
\int {D}_{234}\, W_{34}^{p2}\,,
\quad f(t_1,p)=f(t_1,\vec{x} - \vec{v} (t-t_1),p)\,,
\eeq{newnot1}
\beq
G(t_1,p)=\frac{e^{-\int_{t_1}^{t} dt_2  \sigma n(t_2,\vec{x} - \vec{v} (t-t_2)) v }}{C(x,p)}\,,
\quad C(x,p)=\int_{t_0}^{t} dt_1  
 e^{-\int_{t_1}^{t} dt_2  \sigma n(t_2,\vec{x} - \vec{v} (t-t_2)) v } \, . 
\eeq{newnot2}

The obvious limit in which  MBTE is reduced to BTE 
is a completely homogeneous ST distribution function 
(i.e. no external forces, no boundaries). Another possibility is a 
thermodynamical limit, $\lambda=1/\sigma n\rightarrow 0$, when the exponential factors (\ref{newnot2})
will be reduced to $\sim \delta(t-t_{1,2})$, 
reproducing the BTE after $t_1,t_2$ integrations.

For the BTE the  symmetry of the invariant transition rate lead to the consequence that local conservation
laws can be derived from the original BTE, i.e.
$$\partial_\mu T^{\mu\nu} = 0\quad {\rm and}\quad \partial_\mu N^\mu = 0$$ 
were $T$ and $N$ are given as
momentum-integrals over the single particle PS distribution. 
Although now we have delocalized the equations, the local conservation laws can be still 
derived in the same way.  Let us create a quantity $\Psi_k(x)=a(x)+b(x) p_k^\mu$, which 
is conserved in the binary collisions $12\rightarrow 34$, i.e.  $\Psi_1+\Psi_2=\Psi_3+\Psi_4$. 
Now let us study the quantity $F$:
$$
F=\int \frac{d^3 p_1}{p_1^0} \Psi_1 p_1^\mu \partial_\mu f(x,p_1) = 
$$
\beq
=\int \mathcal{D}_{1234}^{t_1t_2}
 \Psi_1 
\left[f(t_1,p_3) G(t_1,p_3)  \, 
f(t_2,p_4)  G(t_2,p_4) 
- 
f(t_1,p) G(t_1,p_1)  \, 
f(t_2,p_2)  G(t_2,p_2)\right] \,. 
\eeq{ex1} 
Using the  symmetry of the transition rate it is easy to show \cite{LasBook} that 
\beq
F= \int \mathcal{D}_{1234}^{t_1t_2}
f(t_1,p_3) G(t_1,p_3)  \, 
f(t_2,p_4)  G(t_2,p_4) 
\left(\Psi_1(x)+\Psi_2(x)-\Psi_3(x)-\Psi_4(x)\right) = 0\ . 
\eeq{ex2} 
Now if we choose $\Psi_k=q$, where $q$ is a conserved charge, we obtain the charge conservation, 
and if we choose $\Psi_k=p_k^\nu$ we obtain the energy-momentum conservation.

The very essential property of the BTE is the Boltzmann H-theorem. In order to study the entropy
4-current,
$$
S^{\mu}=\int \frac{d^3 p_1}{p_1^0} p_1^\mu f(x,p_1) \left( \log\left(f(x,p_1)\right)-1\right)\,,
$$
we have to choose $\Psi_k(x)=\log\left(f(x,p_k)\right)$, which is not a conserved quantity.
Nevertheless, repeating the same steps as for eq. (\ref{ex2}) we obtain:
$$
S^{\mu}_{,\mu}=\int \mathcal{D}_{1234}^{t_1t_2}
f(t_1,\vec{x} - \vec{v}_3 (t-t_1),p_3) G(t_1,p_3)  \, 
f(t_2,\vec{x} - \vec{v}_4 (t-t_2),p_4)  G(t_2,p_4) \times
$$
\beq
\hfill \times \log\left(\frac{f(x,p_1)f(x,p_2)}{f(x,p_3)f(x,p_4)}\right)\,.
\eeq{Smu2}
The usual way of proving that $S^{\mu}_{,\mu}\ge 0$ does not work for MBTE because of the 
delocalized integral kernel, $f(t_1,\vec{x} - \vec{v}_3 (t-t_1),p_3) G(t_1,p_3)  \, 
f(t_2,\vec{x} - \vec{v}_4 (t-t_2),p_4)  G(t_2,p_4)$. 
The behaviour of the entropy current in MBTE is a subject of the future studies.  Nevertheless, 
the condition of the adiabatic expansion
($S^{\mu}_{,\mu}= 0$) is the
same as for BTE, namely 
$f(x,p_1)f(x,p_2)=f(x,p_3)f(x,p_4)$.

In conclusion, we have shown that the basic assumptions of BTE are not satisfied during the 
FO process and other very fast processes, and, thus, the description should be modified 
if used for such a modeling. We suggest the Modified Boltzmann Transport Equation and 
study some of its properties. Although the use of 
MBTE makes kinetical FO description much more complicated, some simplifications can be 
done \cite{MBTE} and 
the simplified model can be used for the qualitative understanding of the basic FO
features.

\end{document}